\documentstyle[prd,aps,eqsecnum]{revtex}

\input{psfig}

\begin{document}

\title{Exact quantum states for the diagonal Bianchi type
IX model with negative cosmological constant}
\author{Robert Paternoga and Robert Graham}
\address{Fachbereich Physik, Universit\"at-Gesamthochschule
Essen,
45117 Essen, Germany}
\maketitle


\begin{abstract}

Quantum states of the diagonal Bianchi type IX model with
{\em negative} cosmological constant $\Lambda$ are obtained
by transforming the Chern-Simons solution in Ashtekar's
variables to the metric representation. We apply our method
developed earlier for $\Lambda >0$ and obtain five linearly
independent solutions by using the complete set of
topologically inequivalent integration contours in the
required generalized Fourier-transformation. A caustic in
minisuperspace separates two Euclidean regimes at {\em small}
and {\em large} values of the scale parameter from a single
classically interpretable Lorentzian regime in between,
corresponding to the fact that classically these model-Universes
recollapse. Just one particular solution out
of the five we find gives a normalizable probability distribution
on both branches of the caustic. However, in contrast to the
case of positive cosmological constant, this particular
solution neither satisfies the semi-classical no-boundary
condition, nor does the special initial condition it picks out
for $\hbar \to 0$ evolve into a classically interpretable
Universe.

\end{abstract}

\section{Introduction}
Quantum gravity with a non-vanishing cosmological constant
formulated in Ashtekar's spin-connection variables
\cite{4,5,6,7,8} has
interesting physical states given by the exponential of the
Chern-Simons functional \cite{8,9,10} and appropriate
transformations thereof.
In order to elucidate the physical meaning of such states
it is interesting to consider their restrictions to
spatially homogeneous cosmological models. In a recent
paper \cite{0}, henceforth quoted as [I], we considered the
diagonal Bianchi
IX models with positive cosmological constant from this
point of view and found that five linearly independent
physical states in the metric representation could be
derived from the Chern-Simons functional. This set of
solutions was found to be in one to one correspondence
with the set of topologically different integration
contours which exist for the generalized
Fourier-transformation from the Ashtekar-representation to
the metric representation. Due to the positivity of the
cosmological constant the quantum states found in [I]
describe an expanding (or collapsing) classically
interpretable Lorentzian Universe at large scale
parameters. On the other hand, at sufficiently small
scale parameters the action defined by the exponent
of the wavefunction becomes imaginary and can be
associated only with a quantum mechanically allowed
Euclidean Universe. The two "phases" are separated
by a caustic surface in minisuperspace. It was found
that only one of the five linearly independent states
defines a normalizable probability distribution on this
caustic, that this state satisfies the no-boundary
condition of Hartle and Hawking \cite{15,16}
semi-classically for $\hbar \to 0$ (which means on
scales large compared to the Planck scale), and that,
again for $\hbar \to 0$, it picks out an initial condition
which evolves into a classically interpretable Lorentzian
Universe. For details and further literature we refer to [I].

It is now of interest to consider also what happens for
negative cosmological constant, even though it seems very
unlikely that our Universe has $\Lambda <0$ (The
age-problem resulting from recently measured high values
of the Hubble-parameter \cite{11} and the measured large age
of globular
star-clusters and also the observed high density of
galaxies with large red-shifts seem to call for
$\Lambda >0$). Our motivation is rather to try the
method of [I] for a model-Universe which recollapses, i.e.
for which quantum-mechanically
a classically interpretable Lorentzian evolution phase
is bounded for small {\em and} large values of the scale
parameter by Euclidean evolution phases. Therefore, we have
to expect the appearance of two caustic surfaces in the
minsuperspace of these models, one at small 
and the other at large scale parameter.
Is there still a wavefunction, or are there even several,
which give a normalizable probability distribution on
these surfaces, and how are these wavefunctions related
to the no-boundary state?

To answer these questions we apply in section II the method
of [I] to obtain expressions for again five linearly
independent physical states and identify the caustic
surfaces in minisuperspace. In section III we determine
the behavior of the absolute square of the wavefunctions
on the caustic and identify a single physical state which
gives a normalizable probability distribution in this way.
In section IV we summarize our results. Within the narrow
class of models we consider here they seem to rule out,
with high probability, a classically evolving Universe
with $\Lambda <0$.

\section{Quantum states generated by the Chern-Simons
solution}

In this section we want to construct solutions of the
Wheeler-DeWitt equation for the diagonal Bianchi type
IX model with a cosmological constant $\Lambda <0$,

\begin{equation}\label{1.0}
\Biggl \{
\Bigl \lbrack \hbar \partial_{\alpha}\!-\!\Phi_{,\alpha} \Bigr \rbrack
\Bigl \lbrack \hbar \partial_{\alpha}\!+\!\Phi_{,\alpha} \Bigr \rbrack
\!-\!\Bigl \lbrack \hbar \partial_{+}\!-\!\Phi_{,+} \Bigr \rbrack
\Bigl\lbrack \hbar \partial_{+}\!+\!\Phi_{,+} \Bigr \rbrack
\!-\!\Bigl \lbrack \hbar \partial_{-}\!-\!\Phi_{,-} \Bigr \rbrack
\Bigl \lbrack \hbar \partial_{-}\!+\!\Phi_{,-} \Bigr\rbrack
\!+\!3\,(8 \pi)^{2}\Lambda \,e^{6 \alpha} 
\Biggr \}
\,\Psi(\alpha,\beta_{\pm};\Lambda)=0\, ,
\end{equation}
\begin{equation}\label{1.0+}
\mbox{where}\ \ \Phi:=2 \pi\,e^{2 \alpha}\, \mbox{Tr}\,
e^{2 \mbox{{\footnotesize 
$\beta\mbox{\hspace{-1.4 ex}}\beta$}}}  \qquad \mbox{and}\qquad
\mbox{\boldmath $\beta$}=
(\beta_{i j}):=\mbox{diag}\left(\beta_{+}+\sqrt{3}\ 
\beta_{-}, \beta_{+}-
\sqrt{3}\  \beta_{-}, -2\, \beta_{+} \right )\ .
\end{equation}
\\
In this notation $\partial_{+}$ and $\partial_{-}$ denote
derivatives with respect to the variables $\beta_{+}$ and
$\beta_{-}$, respectively. By writing the Wheeler-DeWitt equation
in the form (\ref{1.0})
we have assumed a specific factor-ordering, which is suggested
by a supersymmetric extension of the 
model \cite{20,21,22}. A different factor-ordering is obtained
by considering (\ref{1.0}) with $\Phi$ replaced by $-\Phi$. In
the present paper, as in [I], we will restrict ourselves
to the factor-ordering as in (\ref{1.0}), while a brief comment
on the solutions in the second case $\Phi \to -\Phi$ is given in
appendix A. 

If the expression (\ref{1.0+}) for $\Phi$ is inserted into the
Wheeler-DeWitt equation (\ref{1.0}), the following more explicit
form is obtained
 
\begin{equation}\label{1.1}
\biggl\{
\frac{\hbar^{2}}{3\, \pi^{2}} \left \lbrack  
\frac{\partial^{2}}{\partial \alpha^{2}}
-\frac{\partial^{2}}{\partial \beta_{+}^{\,2}}
-\frac{\partial^{2}}{\partial \beta_{-}^{\,2}} \right
\rbrack 
-\frac{2\, \hbar}{\pi}\, a^{2} 
\ \mbox{Tr}\, e^{2 \mbox{{\footnotesize
$\beta\mbox{\hspace{-1.4 ex}}\beta$}} } +a^{4}
\ \mbox{Tr} \left (e^{4 \mbox{{\footnotesize 
$\beta\mbox{\hspace{-1.4 ex}}\beta$}}}-
2\,e^{-2 \mbox{{\footnotesize 
$\beta\mbox{\hspace{-1.4 ex}}\beta$}}} \right )
+\Lambda\,a^{6} 
\biggr \}\,\Psi(\alpha,\beta_{\pm};\Lambda)=0\ ,
\end{equation}
\\
where we have introduced the mean scale factor
$a:=2\,e^{\alpha}$.
As in the case $\Lambda > 0$, solutions of (\ref{1.1}) can be obtained by a
transformation  to the Ashtekar representation,
where the Chern-Simons functional, restricted to the
Bianchi type IX case, turns out to be an exact solution. 
Two of the Fourier
integrals which occur in the transformation back to the metric
representation can be
carried out analytically without any loss of generality
and afterwards the same one dimensional integral representation
as in [I] is obtained:\footnote{
Here, in contrast to [I], the {\em total} action, including the part 
which effects the similarity transformation between Ashtekar
and metric variables, has been defined as the exponent of the
integrand.} 

\begin{equation}\label{1.2}
\Psi(\kappa, \beta_{\pm}; \lambda)\, \propto   
\int\limits_{{\cal C}} \mbox{d} u\ \exp\left\lbrack\,
\frac{1}{\lambda}\,
f(\sin u;\kappa,\beta_{\pm})\,\right\rbrack\ ,
\end{equation}
\begin{equation}\label{1.3}
\mbox{with} \qquad 
f(z;\kappa,\beta_{\pm}):=2\, \kappa^{2} e^{-2 \beta_{+}}\, 
\frac{z+\cosh \bigl (2 \sqrt{3} \,\beta_{-} \bigr)}
{1-z^{2}}-z^{2}+2\, \kappa e^{2 \beta_{+}} \left (z-\cosh
\bigl (2 \sqrt{3}\, \beta_{-} 
\bigr ) \right)-
\kappa e^{-4 \beta_{+}}\ .
\end{equation}
\\
Here we have introduced the new variable $\kappa$ and parameter
$\lambda$

\begin{equation}\label{1.4}
\kappa:=\frac{1}{12}\,\Lambda\,a^{2}\ ,\qquad
\lambda:=\frac{\hbar \Lambda}{6 \pi}\ ,
\end{equation}
\\
thereby effectively reducing the number of parameters occuring
in (\ref{1.2}), and we shall also make use of the variables $\kappa_{j}$
defined by

\begin{equation}\label{1.4+}
\kappa_{j}:=\kappa\,e^{-\beta_{j}}\ ,
\end{equation}
\\
where the $\beta_{j}$ are the 
entries of the diagonal anisotropy matrix {\boldmath $\beta$}.
The integration contour ${\cal C}$ in the integral-representation 
(\ref{1.2}) can
be chosen quite freely, as long as a sufficiently strong
fall-off for the integrand and its $u$-derivatives at the
borders  $\partial {\cal C}$ of ${\cal C}$ is guaranteed.
The proportionality factor left open in (\ref{1.2}) may
depend on $\lambda$ and will be fixed later. 

\begin{figure}
\begin{center}
\hskip 0 cm
\psfig{figure=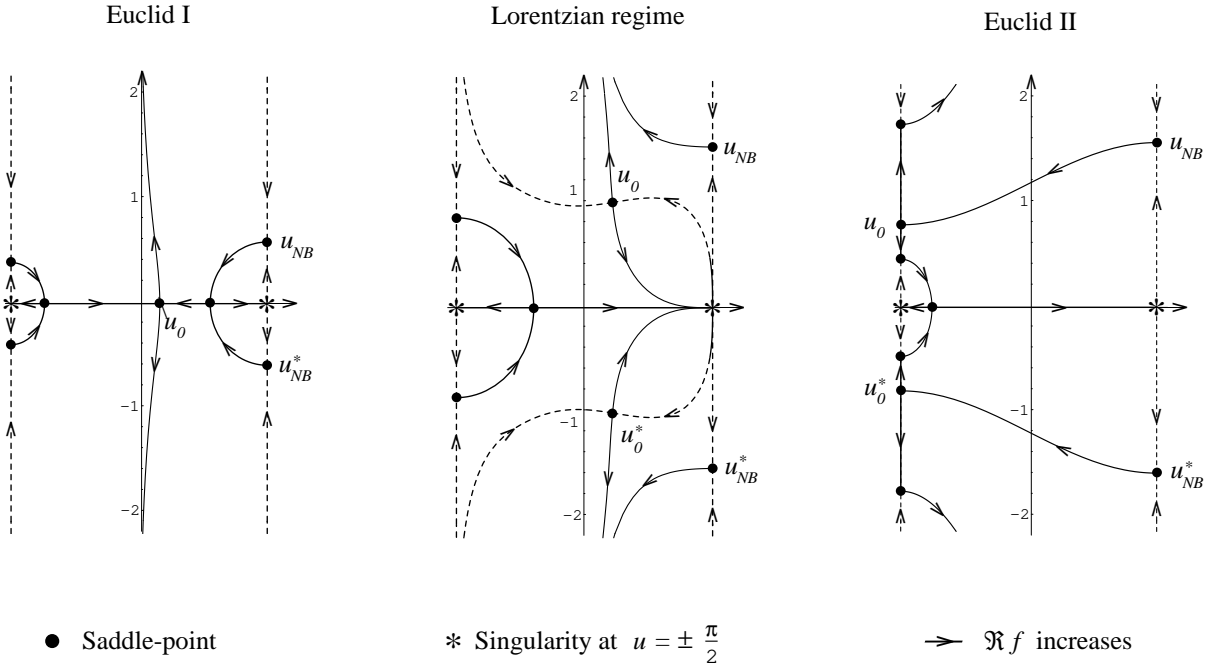,width=16 cm}
\end{center}
\caption{Saddle-points and curves of steepest ascent of
$\Re\,f$ in the complex $u$-plane for $\Lambda <0$. The
picture given in the Lorentzian case only holds for
$\kappa_{3}=\mbox{min}\, \{\kappa_{j}\}$. The remaining
case can easily be constructed by reflecting this figure on
the imaginary axis. The dashed curves come from $- \infty$
with respect to $\Re\,f$ and are given just for
completeness.}
\label{grcsd-}
\end{figure}

While in the case $\Lambda > 0$ the curves of steepest
{\em descent} of $\Re f$ were of interest, now, due to the
different sign of $\Lambda$, the curves of steepest {\em
ascent} lead to suitable integration contours. Moreover,
new possibilities for the location of the saddle-points
occur, which we classify as follows:

\begin{itemize}

\item By choosing $|\kappa |$ sufficiently small at fixed
$\beta_{\pm}$, it is always possible to make all five saddle
points of $f(z)$ lie on the real axis of the complex
$z$-plane, defining the {\em Euclid I}-region of
minisuperspace. Note, however, that the corresponding
points in the $u$-plane of fig.\ref{grcsd-}
(here $u=\mbox{arcsin} z$) are real-valued  only for
$|z| \leq 1$, whereas  real $z$-values with $|z|>1$ are
mapped into complex conjugate pairs of points on the axes  
$\Re \, u=\pm \frac{\pi}{2}$
and periodic repetitions thereof. 

\item Except for the case $\beta_{\pm} =0$, where all five 
saddle-points are on the real $z$-axis, there is the
possibility for two of the saddle-points to become
complex in the $z$-plane, which defines the
{\em Lorentzian regime}.

\item For large values of $| \kappa |$ one always enters
the {\em Euclid II} region, where again all five
saddle-points of $f(z)$ become real-valued.

\end{itemize} 

\noindent
Some typical locations of the saddle-points in these
different regimes of minisuperspace 
and the corresponding curves of steepest ascent are
presented in fig.\ref{grcsd-}.
\vspace{0.5 cm}

\vspace{0.5 cm}

\begin{figure}[t]
\begin{center}
\hskip 0 cm
\psfig{figure=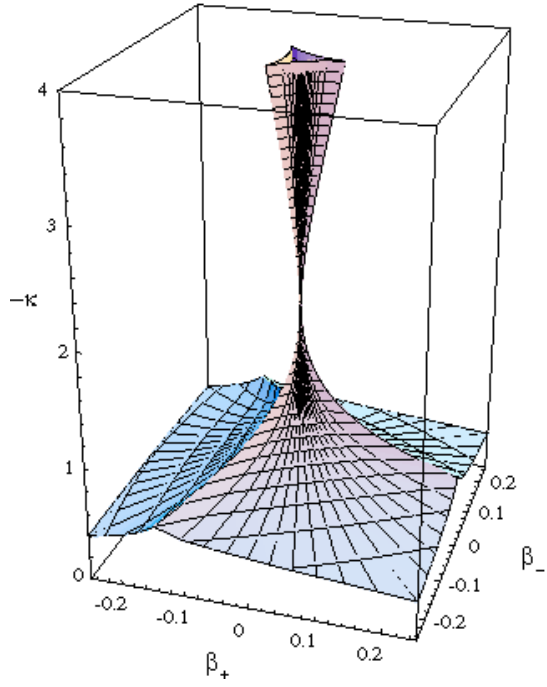,height=9 cm}
\end{center}
\caption{The caustic in minisuperspace for $\Lambda <0$}
\label{grkaustik-}
\end{figure}

\noindent
By passing from one of
these regions to another, a
{\em marginal} situation occurs, where two of the
saddle-points confluate. We will refer to the corresponding
hypersurface in minisuperspace as the {\em caustic}; it
has been calculated and is plotted in fig.\ref{grkaustik-}.
In contrast to the case $\Lambda > 0$ the caustic obtained
here consists of an upper and a lower branch, which are
connected just by a single {\em point} at $ \kappa=-2,\,
\beta_{\pm}=0 $. Furthermore, there are {\em kinks} at
$\beta_{+} >0,\, \beta_{-}=0$ and also at the other half-rays
of the $\beta_{\pm}$-plane, related to the former by the
typical $\beta_{\pm}$-symmetries of diagonal Bianchi IX. 

Obviously, an exactly isotropic Universe $\beta_{\pm}=0$ has to stay
purely Euclidean throughout its evolution. On the other hand,
``large'' Universes with Lorentzian geometry must become
very anisotropic. Apart from the possibility of a negative
cosmological constant very close to zero, which would allow
for large scale parameters even at $|\kappa |$-values of
order one, it seems impossible for the model under
investigation to describe the Universe observed today.
\vspace{0.3 cm}

\noindent
Nevertheless, let us now construct a basis of solutions to
the Wheeler-DeWitt equation (\ref{1.1}) by choosing
topologically independent integration contours ${\cal C}$
in the representation (\ref{1.2}).
\vspace{0.5 cm}

\begin{figure}
\begin{center}
\hskip 0 cm
\psfig{figure=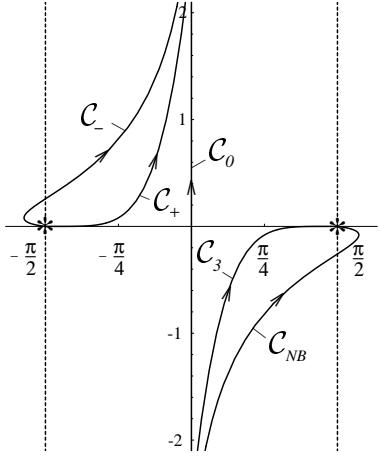,height=6 cm}
\end{center}
\caption{Basis set of integration curves}
\label{grintc}
\end{figure}

\noindent
 Using the curves defined
in fig.\ref{grintc} we introduce the following 
solutions

\begin{equation}\label{1.5}
\Psi_{0}:= \frac{-i\,e^{\mu}}{K_{0}(-\mu)}\,
\int \limits_{\mbox{{\footnotesize${\cal C}_{0}$}}} \mbox{d} u\,
\exp \left \lbrack \frac{1}{\lambda}\,
f(\sin u) \right \rbrack\ , \qquad
\Psi_{\varrho}:=\frac{e^{\mu}}{\pi\,I_{0}(\mu)}\,
\int \limits_{\mbox{{\footnotesize ${\cal C}_{\varrho}
\oplus {\cal C}_{\varrho}^{*}$}}} \mbox{d} u\, \exp \left
\lbrack \frac{1}{\lambda}\,f(\sin u) \right \rbrack\ ,\ \ 
\varrho \,\epsilon\, \{\mbox{{\footnotesize $-$}},
\mbox{{\footnotesize $+$}},3,\mbox{{\scriptsize $NB\,$}}\}
\ ,
\end{equation}
\begin{equation}\label{1.5+}
\mbox{with} \qquad \mu:=\frac{1}{2\,\lambda}\ ,
\end{equation} 
\\
which, by definition, are real-valued and normalized in
accordance with

\begin{equation}\label{1.6}
\Psi_{\varrho}(a=0) \equiv 1\ \ ,\ \ 
\varrho \,\epsilon\, \{0,\mbox{{\footnotesize $-$}},
\mbox{{\footnotesize $+$}},3,\mbox{{\scriptsize $NB\,$}}\}
\ .
\end{equation}
\\
The functions $K_{0}$ and $I_{0}$ occuring in (\ref{1.5})
are the usual modified Bessel functions with index $0$.  
It will be of some advantage to replace the solutions
$\Psi_{+}$ and $\Psi_{-}$ by
      
\begin{equation}\label{1.7}
\Psi_{1}:=\biggl\{
\begin{array}{ccc}
\Psi_{+}\!\!&,&\beta_{-} \geq 0\\
\Psi_{-}\!\!&,&\beta_{-} \leq 0
\end{array}\qquad ,\qquad 
\Psi_{2}:=\biggl\{
\begin{array}{ccc}
\Psi_{+}\!\!&,&\beta_{-}\leq 0\\
\Psi_{-}\!\!&,&\beta_{-} \geq 0
\end{array}\ . 
\end{equation}
\\
Then saddle-point expansions of the integrals (\ref{1.5})
in the limit $\Lambda \to 0$, $a$ and $\beta_{\pm}$ fixed, reveal

\begin{equation}\label{1.8}
\lim_{\Lambda \to 0} \, \Psi_{0}=\Psi_{\mbox{{\tiny $W\!H$}}}^{0}\ ,
\qquad\lim_{\Lambda \to 0} \, \Psi_{\mbox{{\tiny $N\!B$}}}=
\Psi_{\mbox{{\tiny $N\!B$}}}^{0}\ ,\qquad
\lim_{\Lambda \to 0} \, \Psi_{i}=\Psi_{i}^{0}\ ,\ 
i\,\epsilon\,\{1,2,3\}\ ,
\end{equation}
\\
where the upper index ``$0$'' denotes the solutions of the
$\Lambda=0$-model given in [I].

Without proof we mention that 
$\Psi_{i},\,i\,\epsilon\,\{1,2,3\}$, are three asymmetric
solutions which generate each other by cyclic permutations
of the $\kappa_{j}$, so consequently the sum of these
states,

\begin{equation}\label{1.15}
\Psi_{\Sigma}:=\frac{1}{3}\  \sum_{i=1}^{3}\,\Psi_{i}\ ,
\end{equation}
\\
besides $\Psi_{0}$ and $\Psi_{\mbox{{\tiny $N\!B$}}}$, turns out to be
symmetric with respect to arbitrary
$\kappa_{j}$-permutations.

Up to different normalization factors, the asymptotic
behavior in the limit $\kappa \to - \infty$ can immediately
be extracted from the corresponding expansions in [I]
by taking account of the negative sign of $\Lambda$. The
only difficulty lies in the determination of the
saddle-points, which give the dominating contribution to
the different solutions in this limit. A detailed calculation
finally yields  

\begin{equation}\label{1.9}
\Psi_{0}\  \ \  \sim^{^{\!\!\!\!\!\!\!\!\!\!\!\!
\!\mbox{{\scriptsize $\kappa\!
\to \! -\infty$}}}}\,
\frac{\sqrt{\hbar}}{K_{0}\left (-\frac{3 \pi}
{\hbar \Lambda}
\right )}\, \left (-\frac{3}{\Lambda} \right )^{\frac{1}
{4}}\ 
\left ( \frac{a}{2} \right )^{-\frac{3}{2}}\
\exp \left \lbrack - \frac{\pi\,a^{3}}{\hbar}\, \sqrt{
-\frac{\Lambda}{3}} \,\right \rbrack\ ,
\end{equation}
\begin{equation}\label{1.10}
\Psi_{\mbox{{\tiny $N\!B$}}}\  \ \  \sim^{^{\!\!\!\!\!\!\!\!\!\!\!\!
\!\mbox{{\scriptsize $\kappa\!
\to \! -\infty$}}}}\,
\frac{\sqrt{\hbar}}{\pi\,I_{0}\left (\frac{3 \pi}{\hbar
\Lambda}
\right )}\, \left (-\frac{3}{\Lambda} \right )^{
\frac{1}{4}}\ 
\left ( \frac{a}{2} \right )^{-\frac{3}{2}}\
\exp \left \lbrack  + \frac{\pi\,a^{3}}{\hbar}\, \sqrt{
-\frac{\Lambda}{3}} \,\right \rbrack\ ,
\end{equation}
\\
at $\beta_{\pm}=0$, i.e. while $\Psi_{0}$ falls of rapidly for
$a \to \infty$, the wavefunction $\Psi_{\mbox{{\tiny $N\!B$}}}$ is strongly
divergent in the same limit. Moreover, since $\Psi_{\mbox{{\tiny $N\!B$}}}$
{\em always} gets its dominant contribution from the real
saddle-point $z \geq 1$ (corresponding to the points
$u_{\mbox{{\tiny $N\!B$}}}$ and  $u_{\mbox{{\tiny $N\!B$}}}^{*}$ in
the complex $u$-plane of
fig.\ref{grcsd-} via $z=\sin u$), just Euclidean geometries
are described by this state, so we will reject $\Psi_{\mbox{{\tiny $N\!B$}}}$
as a physically relevant solution. Note, however, that it
is the only state which satisfies the {\em no-boundary}
condition in the limit $\hbar \to 0,\,
a \to 0$, hence the name of this wavefunction.

To give the asymptotic behavior in the limit
$\kappa \to - \infty$ for the states
$\Psi_{i},\,i\,\epsilon\,\{1,2,3\}$, it will be helpful
to consider 

\begin{equation}
\Psi^{i}:=\frac{1}{2}\,\left (\Psi_{j}+\Psi_{k} \right )
\ ,\qquad
\varepsilon_{i j k}=1\ ,
\end{equation}
\\
instead. For these solutions the asymptotic expansions

\begin{equation}\label{1.12}
\Psi^{i}\ \ \  \sim^{^{\!\!\!\!\!\!\!\!\!\!\!\!
\!\mbox{{\scriptsize $\kappa\!
\to \! -\infty$}}}}\,  
-\frac{\Psi_{\mbox{{\tiny $W\!H$}}}^{0}}{I_{0}(\mu)}\ 
\sqrt{-\frac{\lambda}{\pi}}\ \frac{2}{\kappa_{i}}\,
\left \{1-2\,\frac{\kappa_{j} \kappa_{k}}{\kappa_{i}^{3}} 
\right \}\,\exp \left \lbrack \frac{1}{\lambda}\, \left (
\kappa_{i}^{2}-2\, \frac{\kappa_{j} \kappa_{k}}{\kappa_{i}}
\right )\,\right \rbrack\ , \  
\varepsilon_{i j k}=1\ ,
\end{equation}
\\
hold, so they fall off very rapidly for $a \to \infty$
(remember the negative sign of $\lambda$ !).

By considering additional asymptotic expansions for large
anisotropy it is possible to show that the four states
$\Psi_{i},\,i\,\epsilon\,\{0,1,2,3\}$, are all normalizable
on minisuperspace in the distribution sense (see [I] for a discussion
of this point for $\Lambda >0$), i.e. so far
we are left with a still four dimensional space of
physically interesting solutions.

However, while in the Lorentzian regime $\Psi_{0}$ receives
saddle-point contributions exclusively from the
saddle-points at {\em complex} $z$ and thus describes a Lorentzian
Universe in this part of minisuperspace, the states
$\Psi_{i},\,i\,\epsilon\,\{1,2,3\}$, 
get additional Euclidean contributions of similar order of
magnitude from {\em real} saddle-points and are therefore
hard to interpret.

The classical trajectories which are generated by $\Psi_{0}$
in the 
semi-classical limit $\hbar \to 0$ {\em in the Lorentzian
regime} can be computed by solving the equations

\begin{equation}\label{1.13}
\frac{\mbox{d} \alpha}{\mbox{d} t}=-\frac{\mbox{d} \Im\,f(z_{0})}
{\mbox{d} \alpha}\ ,\qquad
\frac{\mbox{d} \beta_{\pm}}{\mbox{d} t}= \,\frac{\mbox{d} \Im\,f(z_{0})}
{\mbox{d} \beta_{\pm}}\ ,
\end{equation}
\\
where we have chosen the lapse-function to be
$N=\frac{1}{2}\,\Lambda a^{3}$. While the complex saddle-point $z_{0}$
occuring in (\ref{1.13}) is intended to correspond to the point $u_{0}$ 
of fig.\ref{grcsd-}, the complex conjugate saddle-point $z_{0}^{*}=
\sin u_{0}^{*}$, which describes the time-reversed classical evolution, 
can be considered with the same right. The corresponding second branch of
the classical evolution of the Universe is actually {\em needed}
to define the continuation of a classical trajectory which has reached
the caustic: 
in approaching the caustic the saddle-points $z_{0}$ and
$z_{0}^{*}$ confluate and become real-valued, so that, in accordance
with (\ref{1.13}), the time-derivatives of $\alpha$ and $\beta_{\pm}$ vanish.
To continue such a trajectory in time, the time-reversed version of
(\ref{1.13}) has to be considered. Since
the Universe is "reflected" in this way whenever it meets the caustic,
and since in the generic case the classical trajectories have both of
their endpoints on the caustic, {\em oscillating} Universes are described
by $\Psi_{0}$.

\begin{figure}
\begin{center}
\hskip 0 cm
\psfig{figure=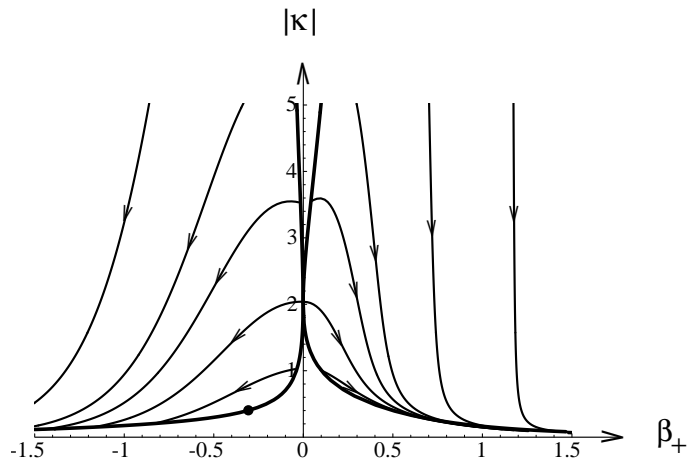,height=6 cm}
\end{center}
\caption{Semi-classical trajectories generated by the
complex saddle-points in the Lorentzian regime. For
simplicity, we have restricted the plot to the plane
$\beta_{-}=0$. The arrows indicate the direction of
increasing time $t$ in eq. (2.18).}
\label{class.traj}
\end{figure}

The numerical results for the classical trajectories which are
obtained in the plane
$\beta_{-}=0$ of the minisuperspace are presented in
fig.\ref{class.traj}.
\vspace{0.3 cm}

\noindent 
We should stress one important  difference in
the pictures which are obtained for the different signs
of $\beta_{+}$:

While for $\beta_{+} >0, \beta_{-}=0$ all trajectories run
to infinite anisotropy (which is, indeed, a peculiarity of the
special $\beta_{\pm}$-direction, corresponding to a kink on the caustic, cf. fig.\ref{grkaustik-}),
in the case $\beta_{+}<0$ the trajectories meet the lower
branch of the caustic again at a finite $\beta_{+}$-value,
representing the general situation. This feature gives rise
to the existence of a special trajectory with coinciding
start- and endpoints, hence describing a Universe that
never really becomes  Lorentzian. The corresponding points
in minisuperspace
can be calculated analytically, requiring
the solution of (\ref{1.13}) to be {\em tangential} to
the caustic, with the result

\begin{equation}\label{1.14}
\kappa=-\sqrt[\!3]{2}\,\left (\frac{2}{5} \right )^{
\frac{4}{3}}\ ,\qquad \beta_{+}+i\,\beta_{-}=\frac{1}{6}
(\ln 5-5 \ln 2 )\,
e^{\mbox{{\footnotesize $\frac{2 \pi i n}{3}$}}}\ ,\qquad
n\,\epsilon\,\{-1,0,1\}\ .
\end{equation}
\\
These points will play an important role in the following
section.

\section{Behavior on the caustic}

Since the classical Lorentzian evolution of the Universe described by
the wavefunctions $\Psi_{i},\,i\,\epsilon\,\{0,1,2,3\}$,
is bounded by the caustic in
minisuperspace, the value of $|\Psi|^{2}_{c}$ {\em on
the caustic} predicted by the different solutions is of
particular interest. In fact $|\Psi|^{2}_{c}$ governs the
realization of the different possible histories of the
Universe and may thus be interpreted as the "initial" value
distribution for the classical evolution.

However, at this stage a new problem arises due to the
different branches of the caustic. Since the semi-classical trajectories
allways can be passed through in both directions, it is impossible 
to distinguish between their start- and endpoints. The distributions of
$|\Psi|^{2}_{c}$ on the upper and lower branch of the caustic
may therefore be considered with the same right, and we
will always discuss them together in the following.

\begin{figure}
\begin{center}
\hskip 0 cm
\psfig{figure=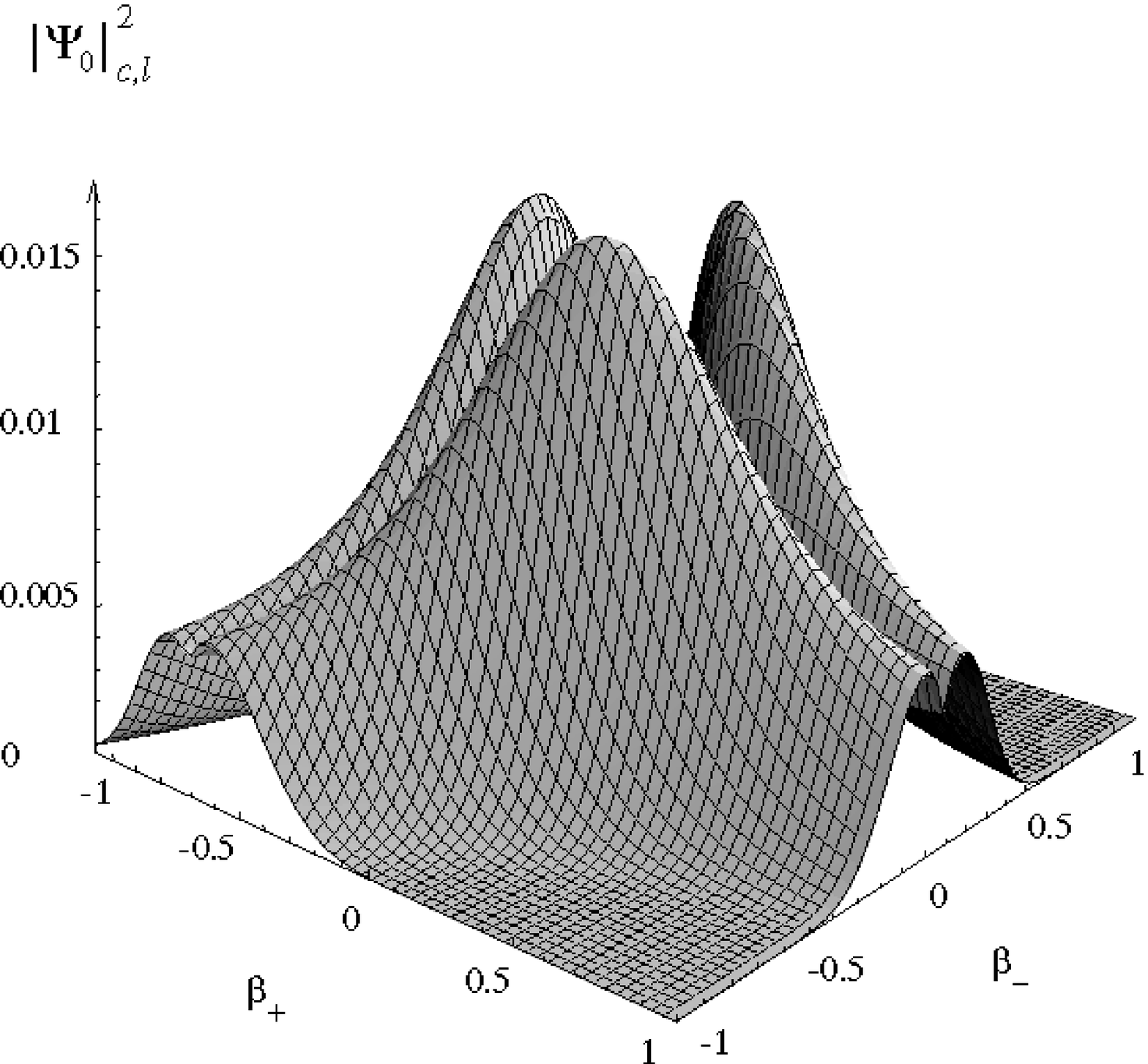,width=7 cm}
\hskip 1.5 cm
\psfig{figure=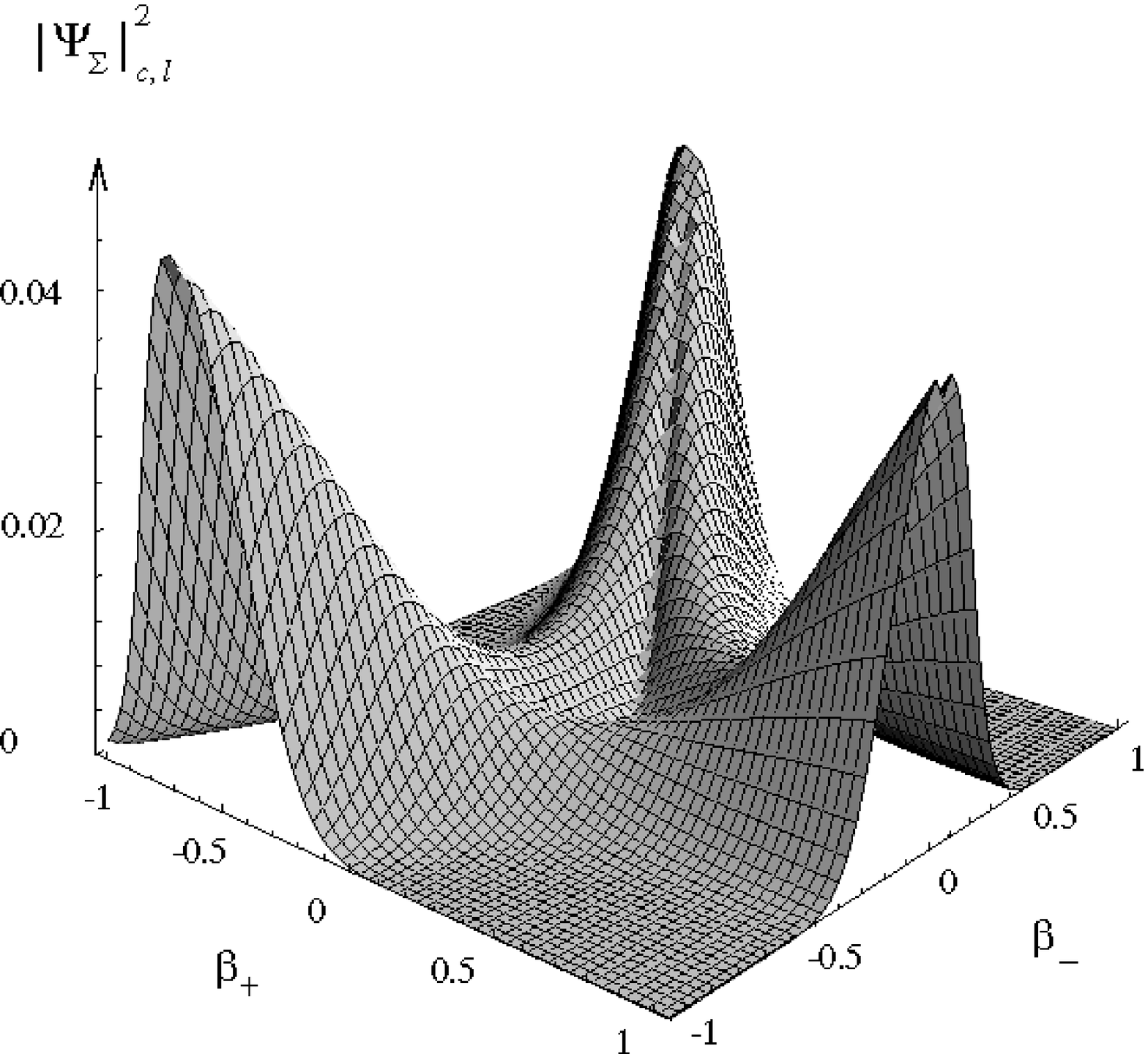,width=7 cm}
\end{center}
\caption{Initial value distributions generated by $\Psi_{0}$
and $\Psi_{\Sigma}$ on the lower caustic($\Lambda=-3,\,
\hbar=2 \pi$).   Like the caustic itself, the distributions
have kinks in some critical $\beta_{\pm}$-directions, which are
partially hidden  in the figures.}
\label{grkaustl}
\end{figure}

\begin{figure}
\begin{center}
\hskip 0 cm
\psfig{figure=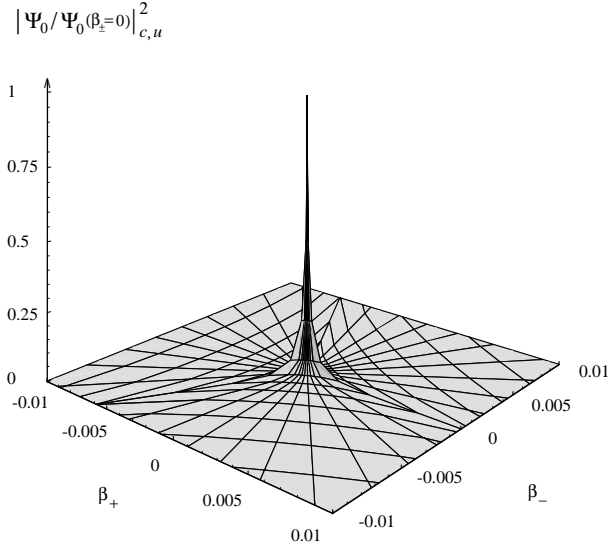,width=8 cm}
\end{center}
\caption{Initial value distribution of $\Psi_{0}$ on the
upper caustic, normalized to unity at $\beta_{\pm} =0$. The
numerical plots obtained for the different wavefunctions
$\Psi_{0}$ and $\Psi_{\Sigma}$ (and thus for $\widehat
\Psi$ defined below) on the upper caustic look very
similar, so we restrict ourselves to a representation
of $|\Psi_{0}|^{2}_{c,u}$. The absolute values taken by
$|\Psi|^{2}_{c,u}$ at $\beta_{\pm} =0$ are given by
$1.45\!\cdot\! 10^{-11},\,3.56\!\cdot\! 10^{-13}$ and
$2.43\!\cdot\! 10^{-11}$ for the wavefunctions
$\Psi_{0},\,\Psi_{\Sigma}$ and  $\widehat
\Psi$, respectively (here again $\Lambda =-3,
\hbar =2 \pi$). Since the lower and upper caustic
coincide at $\beta_{\pm} =0$, it is clear that these values
hold for the distributions on the lower caustic, too.
That is why we suggest to consider the two distributions
obtained on the different branches of the caustic as
analytical continuations of one another through the
isotropic point.}
\label{grkaustu}
\end{figure}

The numerical results obtained for $|\Psi_{0}|^{2}_{c}$ and
$|\Psi_{\Sigma}|^{2}_{c}$  on the lower caustic are given
in fig.\ref{grkaustl}, and fig.\ref{grkaustu} shows the
behavior on the upper caustic, which is very similar for
the two different solutions. In the following the
additional indices "$u$" and "$l$" denote the upper
and lower branch of the caustic, respectively.
\vspace{0.3 cm}

\vspace{0.5 cm}

\begin{figure}
\begin{center}
\hskip 0 cm
\psfig{figure=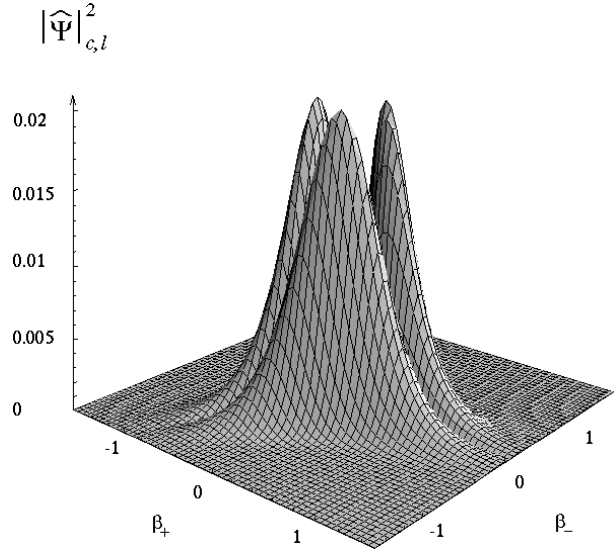,width=8 cm}
\end{center}
\caption{Initial value distribution generated by $\widehat
\Psi$ on the lower caustic ($\Lambda=-3,\,\hbar=2 \pi$).
For the distribution obtained on the upper caustic see
fig.6.}
\label{grkausth}
\end{figure}

\noindent
While on the upper caustic both distributions fall off
rapidly with increasing $\beta_{\pm}$ and may be shown to be
integrable over the $\beta_{\pm}$-plane, there are
$\beta_{\pm}$-directions on the lower caustic, in which 
$|\Psi|^{2}_{c,l}$ approaches a finite value at infinity.
Consequently, the wavefunctions on the lower branch are
{\em not} square-integrable, and hence difficult to
interpret as probability distributions.
Nevertheless, as in the case $\Lambda >0$, one may construct
a new wavefunction as a linear combination of the two symmetric
wavefunctions $\Psi_{0}$ and $\Psi_{\Sigma}$:
By normalizing $\Psi_{0}$ and $\Psi_{\Sigma}$ to approach
unity in the critical $\beta_{\pm}$-directions, the difference
of these new functions is square-integrable on
the {\em full} caustic. To give an explicit expression
for the quantum state obtained in this way, we introduce
the integrals

\begin{displaymath}
{\cal J}_{0}^{(1)}(\nu):=\int \limits_{-\frac{\pi}{4}}^{
+\frac{\pi}{4}} \mbox{d} x \,
e^{-\nu \sin^{4} x}\ ,\qquad
{\cal J}_{0}^{(2)}(\nu):=\int \limits_{-\frac{\pi}{4}}^{
+\frac{\pi}{4}} \mbox{d} x \,
e^{-\nu \cos^{4} x}\ ,
\end{displaymath}
\begin{equation}\label{2.1}
{\cal K}_{0}^{(1)}(\mu):=\int \limits_{0}^{\infty} \mbox{d} x\,
\sin \left (4\,\mu \sinh t \right )\, e^{\mu \cosh 2 t}\ ,
\qquad
{\cal K}_{0}^{(2)}(\mu):=\int \limits_{0}^{\infty} \mbox{d} x\,
\cos \left (4\,\mu \sinh t \right )\, e^{\mu \cosh 2 t}\ ,
\end{equation}
\\
which, as far as we know, have no simple representation in
terms of tabulated functions. Defining now

\begin{equation}\label{2.2}
{\cal Q}(\lambda):=3 \pi\,e^{-3 \mu}\ 
\frac{I_{0}(\mu)}{K_{0}(-\mu)}\ \,
\frac{{\cal K}_{0}^{(2)}(\mu)}{2 {\cal J}_{0}^{(1)}(8 \mu)+
{\cal J}_{0}^{(2)}
(8 \mu)+e^{-3 \mu}\,{\cal K}_{0}^{(1)}(\mu)}\ \  ,\ \ \ 
\mbox{with} \ \ \mu=\frac{1}{2 \lambda}\ ,
\end{equation}
\\
the new state can be written in the form

\begin{equation}\label{2.3}
\widehat \Psi := \frac{\Psi_{0}-{\cal Q}\,\Psi_{\Sigma}}{1-{\cal Q}}
\ ,
\end{equation}
\\
where the overall normalization factor has again been chosen 
to make $\widehat \Psi \equiv 1$ at $a =0$.
The behavior of $\widehat \Psi$ on the caustic has been
computed for fig.\ref{grkausth}. Taking account of the
full distribution, three maxima on the lower branch of
the caustic pick out special initial values for the
classical evolution of the Universe. The general
representation (\ref{1.2}) easily reveals that the
wavefunction becomes arbitrarily sharply concentrated
about these maxima in the limit $\lambda \to 0$, i.e. in
particular in the limit $\hbar \to 0$ at fixed $\Lambda$. Consequently, in
the semi-classical limit there are just three histories
of the Universe which occur with significant probability.

\vspace{.3 cm}
 
\noindent
In the following we shall be interested in the special
points of minisuperspace where the maxima of $| \widehat
\Psi |^{2}_{c,l}$ arise.
Using the saddle-point method for $\lambda \to 0$ the
asymptotic behavior of the integrals defined in (\ref{2.1})
after some calculation yields

\begin{equation}\label{2.4}
{\cal Q} \ \, \to^{^{\!\!\!\!\!\!\!\!\!\!\!
\mbox{{\scriptsize $\lambda\! \to \!0$}}}}\,
\frac{3}{2} \sqrt{\frac{2}{\pi}}\,
\Gamma(\mbox{{\footnotesize $\frac{1}{4}$}})\,
(-\lambda)^{-\frac{1}{4}}\, 
e^{\mbox{{\footnotesize $\frac{3}{\lambda}$}}}\ ,
\end{equation}
\\
and with this result the relation

\begin{equation}\label{2.5}
\widehat \Psi\, \ \, \sim^{^{\!\!\!\!\!\!\!\!\!\!\!
\mbox{{\scriptsize $\hbar\! \to \!0$}}}}\, \Psi_{0}
\end{equation}
\\
can be shown to hold at least on the lower caustic. Since
$\widehat \Psi$ is a real-valued, non-vanishing
wavefunction, the maxima of $|\widehat \Psi|^{2}$
coincide with the maxima of $\widehat \Psi$, and
using (\ref{2.5}) they may also be calculated from
$\Psi_{0}$ in the semi-classical limit. By performing
again a saddle-point expansion for $\lambda \to 0$, now
in the integral representation (\ref{1.5}) of $\Psi_{0}$,
the maxima of $\Psi_{0}$ on the caustic can be calculated
analytically with the result given exactly by
(\ref{1.14}).

Consequently, within the class of solutions considered here,
the {\em only} quantum state that is
square-integrable on the full caustic turns out to
predict a Universe which never becomes
Lorentzian in the classical limit (albeit classical
Lorentzian solutions
of the Bianchi type IX model with negative cosmological
constant actually do exist, cf. fig.\ref{class.traj}).

\section{Conclusion}

In the present paper we constructed exact quantum states for the
diagonal Bianchi type IX model with a negative cosmological constant.
We found
that the method presented in [I] for $\Lambda >0$ is indeed perfectly
applicable to the
model with $\Lambda <0$. As for $\Lambda >0$ it gives
five linearly independent solutions, which are generated
by the Chern-Simons state using topologically different
integration contours in the generalized
Fourier-transformation to the metric representation.
Imposing the condition that the wavefunction be
normalizable on the caustic, just one wavefunction remains,
that turns out to have some nice additional properties:
It is found to be normalizable in minisuperspace in the
distribution sense and it respects the symmetries of
the Bianchi type IX model. However, this state does
{\em not} satisfy the no-boundary condition in the
semi-classical limit in contrast to the case $\Lambda >0$,
and it turns out to predict a Universe that never becomes
Lorentzian, after all. Hence we obtain the result, that,
{\em if} one allows for a non-zero cosmological constant
at all, it should be positive, at least as far as the
Chern-Simons functional related states of the quantized
Bianchi IX model are concerned.

\acknowledgements

Support of this work by the Deutsche Forschungsgemeinschaft
through the Sonderforschungsbereich ``Unordnung und gro{\ss}e
Fluktuationen'' is gratefully acknowledged.

\begin{appendix}

\section{Solutions in a different factor-ordering}

For completeness, and in order to obtain an important argument
for the factor-ordering chosen for the Wheeler-DeWitt equation
(\ref{1.1}), we shall make some comments on a further class of 
solutions, which can again be discussed by using the methods of [I]. 

Considering the Wheeler-DeWitt equation in the form (\ref{1.0}) one
may ask, why we have not chosen a different factor-ordering, which
is obtained by changing $\Phi \to - \Phi$. This choice, of course,
would not have affected the classical Hamiltonian, but the quantum
correction $-\frac{2\, \hbar}{\pi}\,a^{2}\,\mbox{Tr}\,
e^{2 \mbox{{\footnotesize$\beta$}} }$ in (\ref{1.1}) 
would have changed its sign.\footnote{The factor-ordering obtained
in this way corresponds to the ${\cal A}^{+}$-representation
introduced by Kodama in \cite{9},
in contrast to the ${\cal A}^{-}$-representation,
which we have considered up to now.}
Since the coordinate transformation $a \to i\,a, \Lambda
\to -\Lambda$ has exactly the same effect as the above mentioned
change of the factor-ordering, it is possible to discuss the 
solutions of the Wheeler-DeWitt equation in the new
factor-ordering by considering still equation (\ref{1.1}),
but substituting formally $a \to i\,a, \Lambda \to -\Lambda$
in the solutions. In the following it will be more convenient to
use the coordinates $\kappa_{j}$ and $\lambda$ introduced in (\ref{1.4})
and (\ref{1.4+}), which transform like $\kappa_{j} \to \kappa_{j},
\lambda \to -\lambda$ under this substitution. 

It should be clear that the solutions of the Wheeler-DeWitt
equation (\ref{1.1}) are still of the form (\ref{1.2}),
but while we looked at the cases $\kappa >0,
\lambda >0$ in [I] and $\kappa <0, \lambda <0$ in
the present paper, now the remaining sectors
$\kappa >0, \lambda <0$ and $\kappa <0, \lambda >0$
are of interest, which, because of the formal substitution
$\lambda \to -\lambda$ just mentioned, describe
solutions for {\em positive} and {\em negative}
cosmological constant in the {\em new} factor-ordering,
respectively. It is easily checked that the location of
the saddle-points, and therefore the caustic, depends
only on $f(z;\kappa,\beta_{\pm})$ defined 
in eq. (\ref{1.3}). This means that, irrespective
of the sign of $\lambda$, we deal with the caustic of
[I] in the case $\kappa >0$, and with the caustic
fig.\ref{grkaustik-} in the case $\kappa <0$. On
the other hand, it is just the sign of $\lambda$ which
decides whether the integration curves of [I]
(for $\lambda >0$) or of fig.\ref{grintc}
(for $\lambda <0$) give suitable integration contours.

However, constructing the solutions for the new
factor-ordering in this manner and applying the
saddle-point method to the integral representation
(\ref{1.2}) in the limit of large anisotropy
$\beta_{\pm}$, it finally turns out, that
{\em any} solution to the Wheeler-DeWitt
equation in the new factor-ordering
{\em diverges} for $\beta_{\pm} \to \infty$,
at least in some $\beta_{\pm}$-sectors. In
other words, in the new factor-ordering there is
no solution which is normalizable in minisuperspace,
not even in the distribution sense. Furthermore, if 
the behavior of the wavefunctions on the caustic is
considered, actually none of these solutions is
found to be square-integrable with respect to
$\beta_{\pm}$.

Comparing these results with the nice normalizability
properties of the solutions of the Wheeler-DeWitt
equation in the factor-ordering of (\ref{1.0})
presented in [I] and the present paper, we believe
to have a compelling argument to rule out the new
factor-ordering. 
It would be interesting
if this argument could be extended even to
the general, inhomogeneous case of quantum gravity.  

\end{appendix}

\end{document}